\documentclass[9pt, technote, final]{IEEEtran}
\usepackage{amsfonts}
\usepackage{amssymb}
\usepackage{amsmath}
\usepackage{epsf}

	\newcommand\ff{{\mathbb F}}
	\newcommand\rr{{\mathbb R}}

	\newcommand\nn{{\mathbb N}}

\newtheorem{theorem}{\indent Theorem}
\newtheorem{lemma}[theorem]{\indent Lemma}
\newtheorem{proposition}[theorem]{\indent Proposition}

\newcommand{\graph}{{\mathcal{G}}}
\newcommand{\Code}{{\mathbb{C}}}

\newcommand{\code}{{\mathcal{C}}}
\newcommand{\encoder}{{\mathcal{E}}}
\newcommand{\decoder}{{\mathcal{D}}}

\newcommand{\GF}{{\mathrm{GF}}}
\newcommand{\e}{\mathsf{e}}
\newcommand{\blda}{{\mbox{\boldmath $a$}}}
\newcommand{\bldb}{{\mbox{\boldmath $b$}}}
\newcommand{\bldc}{{\mbox{\boldmath $c$}}}

\newcommand{\bldx}{{\mbox{\boldmath $x$}}}
\newcommand{\bldy}{{\mbox{\boldmath $y$}}}
\newcommand{\bldz}{{\mbox{\boldmath $z$}}}
\newcommand{\entropy}{{\mathsf{H}}}

\newcommand{\Prob}{{\mathsf{Prob}}}

\newcommand{\ssf}{{\mathsf{s}}}
\newcommand{\rsf}{{\mathsf{r}}}
\newcommand{\dist}{{\mathsf{d}}}

\newcommand{\bet}{{\mathsf{b}}}
\newcommand{\nsf}{{\mathsf{n}}}
\newcommand{\half}{{\textstyle\frac{1}{2}}}

\newcommand{\qed}{\hspace*{\fill}%
    \vbox{\hrule\hbox{\vrule\squarebox{.667em}\vrule}\hrule}\smallskip}
    \def\squarebox#1{\hbox to #1{\hfill\vbox to #1{\vfill}}}

\title{Decoding of Expander Codes\\at Rates Close to Capacity 
    \thanks{%
    The work of  V. Skachek was supported in part by the 
    DIMACS Special Focus Program in Computational Information Theory and Coding. 
    The material in this paper was presented in part at the 2005 IEEE International Symposium 
    on Information Theory, Adelaide, Australia, September 2005. }
    \thanks{%
    A. Ashikhmin is with the Mathematics of Communicatons Department,
		Bell Laboratories, Lucent Technologies,
		600 Mountain Av., NJ  07974, USA 
    (e-mail: aea@research.bell-labs.com).} 
    \thanks{%
    V. Skachek is with the Computer Science Department, Technion -- Israel Institute of Technology, 
    Haifa 32000, Israel (e-mail: vitalys@cs.technion.ac.il). The work of this author was 
    done in part when he was visiting DIMACS/Bell Laboratories. }
    }

\author{Alexei Ashikhmin \, and \, Vitaly Skachek} 

\begin{document}

\maketitle

\pubid{0000--0000/00\$00.00~\copyright~2006 IEEE}
   
\begin{abstract} 
The decoding error probability of codes is studied as a function of their block length. 
It is shown that the existence of codes with a polynomially small decoding error probability implies the existence
of codes with an exponentially small decoding error probability. Specifically, 
it is assumed that there exists a family of codes of length $N$ and rate 
$R=(1-\varepsilon)C$ ($C$ is a capacity of a binary symmetric channel), 
whose decoding probability decreases  
polynomially in $1/N$. It is shown that if the decoding probability
decreases sufficiently fast, but still only polynomially fast in
$1/N$, then there exists another such family of codes
whose decoding error probability decreases exponentially fast in $N$. 
Moreover, if the decoding
time complexity of the assumed family of codes is polynomial in $N$ and $1/\varepsilon$, then the decoding 
time complexity of the presented family is linear in $N$ and polynomial in $1/\varepsilon$. 
These codes are compared to the recently presented codes of Barg and Z\'emor, ``Error Exponents of 
Expander Codes,'' \emph{IEEE Trans. Inform. Theory,} 2002,
and ``Concatenated Codes: Serial and Parallel,'' \emph{IEEE Trans. Inform. Theory,} 2005. 
It is shown that the latter families can not be tuned to have exponentially decaying (in $N$) 
error probability, and at the same time to have decoding time complexity  
linear in $N$ and polynomial in $1/\varepsilon$.


\end{abstract}

\section{Introduction}
\label{sec-intro}

\pubidadjcol

A classical work of Shannon states that reliable communications over
a communication channel can be achieved for all information rates which are less than 
the certain threshold rate, capacity, which is a function of the channel 
characteristics. Codes and decoding algorithms that attain the channel 
capacity were extensively studied over the last decades. For such codes with respective
decoding algorithms, at rates less than the capacity, the probability of decoding error approaches 
zero, as the code length grows. 

Fastness of decrease of the decoding error probability as a function of the code length, $N$,
is a characteristic of capacity-approaching codes, which 
was widely studied for many code families. However, 
this probability depends also on ratio between the channel capacity and
an actual code rate. Namely, let the code rate be $R = (1 - \varepsilon)C$, 
where $C$ is the channel capacity. It is an interesting question to ask is
how the decoding error probability depends on $\varepsilon$.  

Another characteristic of (decoding algorithms of) codes is a time complexity of decoding. 
As of yet, there are known families of 
capacity-achieving codes (over various channels) with decoding algorithm time 
complexity only linear in $N$. However, one might look onto 
the decoding time complexity of code families in terms of $\varepsilon$. 
In the next two paragraphs we discuss these characteristics for two code families. 

It is known that LDPC-type codes can
attain a capacity of a binary erasure channel (BEC), the reader can refer to
~\cite{LMSS},~\cite{Shokrollahi},~\cite{Urbanke2}. It is generally believed that
LDPC-type codes can approach capacity of a variety of other communication
channels. However, it is also believed that the  
decoding error probability decreases only polynomially with the code length. 
As to the decoding time complexity, it was conjectured in~\cite{McEliece} that per-bit complexity of message-passing
decoding (e.g. \cite{Gallager-monograph}, \cite{Urbanke1}) of LDPC
or irregular repeat accumulative (IRA) codes over any `typical' channel is $O\left(\log \frac{1}{\pi}\right) + O\left(\frac{1}{\varepsilon} \log \frac{1}{\varepsilon}\right)$, where $\pi$ is a decoded error probability.
Lately, for LDPC-type codes with message-passing decoding 
over the BEC, the time complexity was shown to be linear in a code length and sub-linear in $1/\varepsilon$. 
More specifically, 
it was shown in~\cite{LMSS} and~\cite{Shokrollahi} that the decoding complexity per bit for some sub-families
of LDPC-type codes behaves as $O(\log (1/\varepsilon))$. Recently, in~\cite{Sason-Urbanke}, IRA codes with bounded decoding complexity per bit were constructed. 

\pubidadjcol

In contrast, modifications of expander codes presented in~\cite{Zemor02}, \cite{Zemor-DIMACS},~\cite{Zemor03},~\cite{Roth-Skachek-2004},~\cite{Roth-Skachek-2004-full}
also attain the capacity of the memoryless $q$-ary 
symmetric channel, and the error probability decreases exponentially with the
code length. Several recent works were devoted to analysis of fraction of errors that expander codes can correct 
(e.g.~\cite{Feldman-fraction},~\cite{SipSpielman},~\cite{Skachek-Roth-2003},~\cite{Zemor01}) 
and their rate-distance trade-offs (see~\cite{Zemor03},~\cite{GI3},~\cite{Roth-Skachek-2004-full}).  
While it is well known that there are decoders for expander codes having linear-time 
(in the code length) complexity, the dependence of this complexity on $1/\varepsilon$ 
was not studied. In the present work, we aim at studying this dependence.
We investigate time complexity of decoding algorithms of expander 
codes in terms of $\varepsilon$, in particular for the codes in \cite{Zemor02},~\cite{Zemor03}. 
We show that these specific codes have time complexity that is exponential in $1/\varepsilon^2$. 

In this work, we study capacity-achieving codes over a binary symmetric 
channel (BSC). We show that if there exists a family of codes $\code_{in}$ of length $N$ and rate 
$R = (1 - \varepsilon)C$ ($C$ is a BSC capacity), with the decoding probability vanishing 
inverse polynomially in $N$ and $\varepsilon$ (under conditions of our theorem), then 
there exists another such family of codes $\Code_{cont}$
with the decoding error probability vanishing exponentially in $N$. 
Moreover, if the decoding
time complexity of the codes $\code_{in}$ is polynomial in $N$ and $1/\varepsilon$, then the decoding 
time complexity of the codes $\Code_{cont}$ is linear in $N$ and polynomial in $1/\varepsilon$. 

The structure of this paper is as follows. In Section~\ref{sec:preliminaries}, we describe 
the basic ingredients in our construction. The main result of our paper appears in 
Section~\ref{sec:main-results}: we present a sufficient condition for existence of a family
of codes with the decoding error probability vanishing exponentially fast. We also analyze 
the decoding time complexity of the presented codes. Finally, in Sections~\ref{sec:analysis-zemor02}
and~\ref{sec:analysis-zemor03}, we show that the codes in~\cite{Zemor02},~\cite{Zemor03} 
with their respective algorithms cannot be tuned to have decoding error probability that 
decreases exponentially fast (in terms of $N$), while the 
respective decoding algorithms have time complexity linear in $N$ and 
polynomial in $1/\varepsilon$.

\section{Preliminaries}
\label{sec:preliminaries}

\subsection{Capacity-achieving codes with fast decoding} 
\label{sec:assumed-code}

In this subsection we assume existence of some (family of) linear code $\code_{in}$, 
which achieves the capacity $C$ 
of the BSC, and which has fast decoding algorithm. We denote its rate $R_{in} = (1 - \varepsilon)C$, 
and its length $n_{in}$ (constant for a fixed $\varepsilon$). Below, we discuss the parameters of this 
code. 
$\,$
\vspace{1ex}
\newline
{\bf Decoding complexity:} 
we assume that the decoding complexity of $\code_{in}$ over the BSC is given by 
\begin{equation}
O\left(n_{in}^{\ssf} \cdot \frac{1}{\varepsilon^\rsf}\right) \; ,
\label{eq:LDPC-complexity}
\end{equation}
where $\ssf, \rsf \ge 1$ are some constants. Let $\decoder_{in}$ be a decoder that have a time complexity 
as in~(\ref{eq:LDPC-complexity}). 

Based on the results in~\cite{LMSS},~\cite{Shokrollahi},~\cite{Sason-Urbanke}, several LDPC-type code families (with 
respective message-passing decoding algorithms) do have such decoding complexity over the BEC (for $\ssf = 1$). 
There are no such results known for the BSC, although in the light of the surveyed works, this assumption 
sounds reasonable for LDPC-type codes over the BSC.  
\newline
{\bf Decoding error probability:} as of yet, there are no satisfying 
results on asymptotical behavior 
of the decoding error probability of LDPC-type codes 
over the binary erasure channel under the message-passing 
decoding, for rates near capacity of the BEC. The behavior of the
decoding error probability of LDPC-type codes over other channels is even
less investigated. In this work, we obtain a sufficient condition on
the probability of the decoding error
$\Prob_e(\code_{in})$ of the decoder $\decoder_{in}$ (for the $\code_{in}$) to guarantee 
the existence of a code with an exponentially-fast decreasing error probability. 
\newline
{\bf Note:} the results presented in the sequel are valid for any code $\code_{in}$ 
whose decoding time complexity 
and error probability are as stated above. However, LDPC-type codes are very promising candidates 
to meet these conditions, and in fact we do not see any other candidate at the present moment. 
Since there is no such candidate, it makes sense to speak about LDPC-type codes in this context. 

\subsection{Nearly-MDS expander codes} 
\label{sec:expander-code}

In this section, we consider linear-time decodable codes of rate $1 - \epsilon$ (for small $\epsilon>0$) 
that can correct a fraction $\vartheta \epsilon^\bet$ of errors, where $\vartheta > 0, \bet>0$ are constants. 
There are several code families known to date that can be shown to have the above property,
and at the same time allow a linear-time (in a code length) decoding. In this connection, the reader
can refer to~\cite{Zemor02},~\cite{Zemor03},~\cite{GI3},~\cite{SipSpielman},~\cite{Zemor01}. 
However, as of yet, the codes in~\cite{Roth-Skachek-2004},~\cite{Roth-Skachek-2004-full}
have the best relations between their rate, distance and alphabet size
among all known expander-based linear-time decodable codes. 
Moreover, unlike the codes in~\cite{Roth-Skachek-2004},~\cite{Roth-Skachek-2004-full}, 
not all aforementioned codes have decoding time complexity, which is polynomial in $1/\epsilon$. 

Below, we recall the construction in~\cite{Roth-Skachek-2004},~\cite{Roth-Skachek-2004-full}. 
Let $\graph = (A:B, E)$ be 
a bipartite $\Delta$-regular undirected connected graph with
a vertex set $V = A \cup B$ such that
$A \cap B = \emptyset$ and $|A| = |B| = n$,
and an edge set $E$ of size $N = \Delta n$ such that
every edge in $E$ has one endpoint in $A$ and one endpoint in $B$. 
For every vertex $u \in V$, denote by $E(u)$ the set of
edges incident with $u$, and assume some ordering on $E(u)$, 
for every $u \in V$. Let $\ff = \GF(q)$ be some finite field, and
$q \ge \Delta$. 

Take $\code_A$ and $\code_B$ to be
Generalized Reed-Solomon codes with parameters $[\Delta, r_A \Delta,\delta_A \Delta]$ and 
$[\Delta,r_B \Delta,\delta_B \Delta]$ over $\ff$, respectively. (We use notation $[n, k ,d]$ 
for a linear code of length $n$, dimension $k$, and minimum distance $d$.)     
We define the code $\Code = (\graph, \code_A : \code_B)$ as in~\cite{Roth-Skachek-2004-full}, namely
\begin{eqnarray}
\Code & = & \left\{ \bldc \in \ff^N \,:\,
(\bldc)_{E(u)} \in \code_A \;
\mbox{for every} \; u \in A \; \nonumber \right. \\
& & \quad \left. \mbox{ and } \; 
(\bldc)_{E(u)} \in \code_B
\mbox{ for every } \; u \in B \right\} \; ,
\label{eq:define_c}
\end{eqnarray}
where $(\bldx)_{E(u)}$ denotes the sub-word of $\bldx = (x_e)_{e \in E} \in \ff^N$ 
that is indexed by $E(u)$. The produced code $\Code$ is a linear code of 
length $N$ over $\ff$.

Let $\Phi$ denote the alphabet $\ff^{r_A \Delta}$. Taking some linear one-to-one 
mapping $\encoder_A : \Phi \rightarrow \code_A$ over $\ff$, and
the mapping $\psi : \Code \rightarrow \Phi^n$ given by
\[
\psi(\bldc) = 
\left(
\encoder_A^{-1} \left( (\bldc)_{E(u)} \right)
\right)_{u \in A}
 \; ,
\quad \bldc \in \Code \; ,
\]
the authors of~\cite{Roth-Skachek-2004-full} 
define the code $\Code_\Phi$ of length $n$ over $\Phi$ by 
\begin{equation*}
\Code_\Phi = \left\{ \psi(\bldc) \; : \; \bldc \in \Code \right\} \; .
\end{equation*}

{\it Definition.} An infinite sequence $\{a_i\}_{i=1}^\infty$,  
$a_i \stackrel{i \rightarrow \infty}{\longrightarrow} +\infty$, $a_i \in \rr$, is called 
\emph{a dense sequence of values} if $a_1 \le 100$ and $a_{i+1} - a_i = o(a_i)$ (for $i \rightarrow \infty$). 
(The number $100$ is a large absolute constant, the condition $a_1 \le 100$ ensures that 
not all elements in the sequence are exponentially large.) 

Let $\lambda_\graph$ be the second largest eigenvalue of
the adjacency matrix of $\graph$
and denote by $\gamma_\graph$ the value $\lambda_\graph / \Delta$.
When $\graph$ is taken from a family of $\Delta$-regular bipartite Ramanujan graphs 
(e.g.~\cite{LPS},~\cite{Margulis}), we have 
\begin{equation}
\lambda_\graph \le 2 \sqrt{\Delta - 1} \; .
\label{eq:ramanujan}
\end{equation}
There are explicit constructions for such $\Delta$-regular Ramanujan graph families
for dense sequences of values $\Delta$ (\cite{LPS},~\cite{Margulis}). 

It was shown in~\cite{Roth-Skachek-2004-full}, that the code $\Code_\Phi$ has  
the relative minimum distance 
\begin{equation}
\delta_\Phi \ge \frac{\delta_B - \gamma_\graph \sqrt{\delta_B/\delta_A}}{1 - \gamma_\graph} \; .
\label{eq:expander-distance}
\end{equation}
It is also known that the rate of $\Code_\Phi$ is  
\[
R_\Phi \ge r_A + r_B - 1 \; . 
\]
The linear-time decoding algorithm $\decoder_\Phi$ in Figure~\ref{fig:decoder} 
was proposed in~\cite{Roth-Skachek-2004-full}. It corrects any
pattern of $\mu$ errors and $\rho$ erasures such that $\mu + \half \rho < \beta n$, 
where $\beta$ is given by
\begin{equation}
\label{eq:beta}
\beta = \frac{(\delta_B/2) - \gamma_\graph \sqrt{\delta_B / \delta_A}}{1 - \gamma_\graph} \; .
\end{equation}
The number of iterations $m$ in the algorithm was established 
in~\cite{Roth-Skachek-2004-full} such that $m=O(\log n)$.  
The notation ``?'' is used for erasures, and the notations $\decoder_A$ and $\decoder_B$ are used 
for decoders of the codes $\code_A$ and $\code_B$, respectively.  

\begin{figure}[hbt]
\makebox[0in]{}\hrulefill\makebox[0in]{}
{\normalsize
\begin{description}
\item[\underline{\bfseries Input:}]
$\;$ received word $\bldy = (\bldy_u)_{u \in A}$
in $( \Phi \cup \{ ? \} )^n$.
\item[\underline{\bfseries For}] $u \in A$ \underline{\textbf{do}}
$\quad (\bldz)_{\!\scriptscriptstyle E(u)} \leftarrow
\left\{ \begin{array}{ll}
\encoder_A (\bldy_u) & \mbox{if } \bldy_u \in \Phi \\
??\cdots? & \mbox{if } \bldy_u = ? \\ 
\end{array}\right.$ .
\item[\underline{\bfseries For}] $i \leftarrow 1,2,\ldots,m$
\underline{{\bfseries do}} \{
\begin{description}
\item[\underline{\bfseries If}]
$i$ is even \underline{{\bfseries then}} $X \equiv A$,
$\decoder \equiv \decoder_A$, \\
\underline{\bf{else}} $X \equiv B$, $\decoder \equiv \decoder_B$.
\item[\underline{\bfseries For}]
$u \in X$ \underline{{\bfseries do}}
$(\bldz)_{\!\scriptscriptstyle E(u)}
\leftarrow \decoder((\bldz)_{\!\scriptscriptstyle E(u)})$.
\end{description} 
\item[] \}
\item[\underline{\bfseries Output:}] $\;\;\;$ $\psi(\bldz)$ if $\bldz \in \Code$ (and declare `error' otherwise).
\end{description}
}
\makebox[0in]{}\hrulefill\makebox[0in]{}
\caption{Decoder $\decoder_\Phi$ of Roth and Skachek for the code $\Code_\Phi$.}
\label{fig:decoder}
\end{figure}

The proof in~\cite{Roth-Skachek-2004-full} requires that the
decoder $\decoder_A$ is a mapping $\ff^\Delta \rightarrow \code_A$
that recovers correctly any pattern of less than $\delta_A \Delta / 2$
errors over $\ff$,
and the decoder $\decoder_B$ is a mapping 
$(\ff \cup \{ ? \})^\Delta \rightarrow \code_B$
that recovers correctly any pattern of $\theta$ errors
and $\nu$ erasures, provided that $2 \theta + \nu < \delta_B \Delta$. 
The decoders $\decoder_A$ and $\decoder_B$ are polynomial-time, for example 
Berlekamp-Massey decoder can be used for both of them. 
It can be implemented then in $O(\Delta^2)$ time (or less).  

In the next proposition, we show that the parameters of 
the codes in~\cite{Roth-Skachek-2004-full} of rate $1 - \epsilon$
can be tuned to correct $\vartheta \epsilon$ errors for a constant $\vartheta > 0$. 

\begin{proposition}
For any $\epsilon \in (0, 1)$, and for a sequence of alphabets $\{ \Phi_i \}_{i=1}^\infty$ such that the sequence
$\{\log_2 |\Phi_i|\}_{i=1}^\infty$ is dense, 
the codes $\Code_\Phi$ (as above) of rate $R_\Phi \ge 1-\epsilon$ 
(with decoder $\decoder_\Phi$) can correct a fraction
$\vartheta \epsilon$ of errors, where $\vartheta > 0$ is some constant. 
\label{prop:parameter-selection}
\end{proposition}

{\bf Proof.} There is a dense sequence of values $\Delta \in \{\Delta_i\}_{i=1}^\infty$ such that
there exists a family of $\Delta$-regular bipartite Ramanujan graphs $\graph$ 
(see~\cite{LPS},~\cite{Margulis}). For any such value $\Delta$, 
we can take both codes $\code_A$ and $\code_B$ to be GRS codes of length $\Delta$ over alphabet of size $\Delta$, 
rate $r_A = r_B = 1 - \epsilon/2$ and relative minimum distance $\delta_A = \delta_B = \epsilon/2$. Consider 
a code $\Code_\Phi$ defined with respect to these $\code_A$ and $\code_B$. The rate $R_\Phi$ of $\Code_\Phi$ 
satisfies $R_\Phi \ge r_A + r_B - 1 = 1 - \epsilon$. From~(\ref{eq:beta}),
the fraction of errors that the decoder $\decoder_\Phi$ can correct is given by
\begin{eqnarray*}
\beta & = & \frac{\delta_B/2 - \gamma_\graph \sqrt{\delta_B/\delta_A}}{1 - \gamma_\graph} \\
& \ge & \epsilon / 4 - \gamma_\graph \\
& = & \epsilon / 4 - 2 \sqrt{\Delta-1}/\Delta \\
& \ge & \epsilon / 4 - 2 / \sqrt{\Delta} \; . 
\end{eqnarray*}

Take any $\Delta$ such that $\Delta > (16 / \epsilon)^2$: for such $\Delta$, 
\[
\beta > \vartheta \epsilon \; , \mbox{ where } \vartheta = 1/8 \; . 
\]

Next, we observe that $|\Phi_i| = \Delta_i^{\Delta_i r_A}$. 
Based on the density of $\{ \Delta_i \}_{i=1}^\infty$, 
we show the density of the sequence $\{\log_2 |\Phi_i|\}_{i=1}^\infty$. Indeed,
for any $i \in \nn$,
\begin{eqnarray*}
\lefteqn{ 
\lim_{i \rightarrow \infty} \frac{\log_2 |\Phi_{i+1}| - \log_2 |\Phi_{i}|}{\log_2 |\Phi_{i}|}}
\makebox[1ex]{} && \\
& = & \lim_{i \rightarrow \infty} 
\frac{\Delta_{i+1} \log_2 \Delta_{i+1} - \Delta_i \log_2 \Delta_{i}}{\Delta_i \log_2 \Delta_{i}} \\
& = & \lim_{i \rightarrow \infty} \Bigg(
\frac{\Delta_{i+1} \log_2 \Delta_{i+1}}{\Delta_i \log_2 \Delta_{i}}\Bigg) - 1 \\
& = & \lim_{i \rightarrow \infty} \Bigg(
\frac{\Delta_{i} + o(\Delta_{i})}{\Delta_i} \cdot 
\frac{\log_2 (\Delta_{i} + o(\Delta_{i})) }{ \log_2 \Delta_{i}} \Bigg) - 1 \\
& = & 1 \; - \; 1 \; = \; 0 \; . 
\end{eqnarray*}
Finally, from~\cite{LPS} and~\cite{Margulis}, 
$\Delta_1$ can be taken small enough, such that $\log_2|\Phi_1| < 100$, as required.  
\qed

\subsection{Concatenated codes} 
\label{sec:concatenated-code}

In this subsection, we revisit the definition of concatenated codes. 
The following ingredients will be used: 
\begin{itemize}
\item
A linear $[\nsf_{in}, k_{in}{=}R_{in} \nsf_{in}]$ code $\code_{in}$ over $\ff$ (inner code).
\item
A linear code $\Code_\Phi$ of length $n$ and rate $R_\Phi$ over $\Phi = \ff^{k_{in}}$ (outer code). 
\item
A linear one-to-one mapping $\encoder_0 \; : \; \Phi \rightarrow \code_{in}$. 
\end{itemize}

The respective concatenated code $\Code_{cont}$ is defined as 
\begin{eqnarray*}
\Code_{cont} = \Big\{ ( \bldc_1 | \bldc_2 | \cdots | \bldc_n ) \in \ff^{n \cdot \nsf_{in}} 
: \bldc_i = \encoder_0(\Xi_i) \; , 
\\ \mbox{ for } i \in 1, 2, \cdots, n, \; 
\mbox{ and } ( \Xi_1 \Xi_2 \cdots \Xi_n) \in \Code_\Phi \Big\} \; . 
\end{eqnarray*}
The rate of $\Code_{cont}$ is known to be $R_{cont} = R_{in} \cdot R_\Phi$.

Let $\decoder_{in} : \ff^{\nsf_{in}} \rightarrow \code_{in}$ and 
$\decoder_\Phi : \Phi^n \rightarrow \Code_\Phi$ be decoders 
for the codes $\code_{in}$ and $\Code_\Phi$, respectively. 
A simple decoder $\decoder_{cont}$ 
for the code $\Code_{cont}$ is presented in Figure~\ref{fig:decoder-concatenated}. There exist 
more advanced decoders for the code $\Code_{cont}$ (e.g. GMD decoding,~\cite{ForneyMonograph}) 
that can correct more errors, but we consider the decoder $\decoder_{cont}$ due to its
simplicity. 
 
\begin{figure}[hbt]
\makebox[0in]{}\hrulefill\makebox[0in]{}
{\normalsize
\begin{description}
\item[\underline{\bfseries Input:}]
$\,$ received word $\bldy = (y_1 \, y_2 \, \cdots \, y_{n \cdot \nsf_{in}})$ in $\ff^{n \cdot \nsf_{in}}$.
\item[\underline{\bfseries For}] $i \in 1, 2, \cdots, n$ \underline{\textbf{do}} \\
{\phantom{aaa}} $u_i \leftarrow \encoder_0^{-1} \left(\decoder_{in} \left( \; 
(y_{j + (i-1) \cdot \nsf_{in}})_{j=1}^{\nsf_{in}} \; \right) \right)$. 
\item[\underline{\bfseries Let}] $(z_1 z_2 \cdots z_n) \leftarrow \decoder_\Phi\left((u_1 u_2 \cdots u_n)\right)$. 
\item[\underline{\bfseries Output:}] 
${\phantom{ai}} \left( \encoder_0(z_1) | \encoder_0(z_2) | \cdots | \encoder_0(z_n) \right)$.
\end{description} 
}
\makebox[0in]{}\hrulefill\makebox[0in]{}
\caption{Decoder $\decoder_{cont}$ for the code $\Code_{cont}$.}
\label{fig:decoder-concatenated}
\end{figure}

\section{Main results}
\label{sec:main-results}

\subsection{General settings}
\label{sec:main-settings}

Consider a memoryless binary symmetric channel with cross\-over 
probability $p$. Its capacity is given by $C = 1 - \entropy_2(p)$,
where $\entropy_2(\chi) = -\chi \log_2 \chi - (1 - \chi) \log_2 (1 - \chi)$ 
is the binary entropy function. Let 
$R = C (1 - \varepsilon)$ be a design rate.

Take $\ff$ to be $\GF(q)$, $q = 2^\ell$, $\ell \in \nn$. 
Let $\code_{in}$ be a binary code of length $n_{in}$ assumed in Section~\ref{sec:assumed-code}.
It can also be seen as an additive linear code of length $\nsf_{in} = n_{in}/\ell$ over $\ff$. 
Let $\Code_\Phi$ be a linear code of length $n$ and rate $R_\Phi$ 
over an alphabet $\Phi = \ff^{R_{in} \nsf_{in}}$. 
Pick some linear one-to-one mapping $\encoder_0 \; : \; \Phi \rightarrow \code_{in}$. 
Let $\Code_{cont}$ be a code, corresponding to a concatenation of the code 
$\code_{in}$ (as an inner code) with the code $\Code_\Phi$ (as an outer code),
as defined in Section~\ref{sec:concatenated-code}.
Suppose $R_{cont} \ge R$ is a rate of the (binary) code $\Code_{cont}$ and $N_{cont} = n \cdot n_{in}$ is its length.
Denote by $\Prob_e(\Code_{cont})$ its error probability, under the decoding by $\decoder_{cont}$.

The following lemma is based on the result in~\cite[Chapter 4.2]{ForneyMonograph}. 
\begin{lemma}
The error probability of the code $\Code_{cont}$ (as defined in this section)
under the decoding by $\decoder_{cont}$, 
when the error probability of the decoder $\decoder_{in}$ for the code $\code_{in}$ is $\Prob_e(\code_{in})$, 
and the decoder $\decoder_\Phi$ corrects any pattern of less than $\beta n$ errors, is bounded by
\[  
\Prob_e(\Code_{cont}) \le \exp \{-n \cdot E \} =  \exp \left\{- N_{cont} \cdot \frac{E}{n_{in}} \right\} \; ,
\]
where $E$ is a constant given by
\begin{eqnarray}
E & = & -\beta \ln \left(\Prob_e(\code_{in})\right) 
- \left(1 - \beta \right) \ln \left(1 - \Prob_e(\code_{in})\right) \nonumber \\
&& + \beta \ln \left( \beta \right) + 
\left(1 - \beta \right) \ln \left(1 - \beta \right) \; . 
\label{expr:forney}
\end{eqnarray}
If a right-hand side of~(\ref{expr:forney}) is negative, we assume that $E$ is zero. 
\label{lemma:forney}
\end{lemma}
 
The proof of this lemma appears in Appendix A.

{\bf Remark.} It is possible to improve an error exponent by a constant factor if allowing 
the decoder for the code $\code_{in}$ to put out an ``erasure''
message in a case of unreliable decoding of the code $\code_{in}$. 
See~\cite[Chapter 4.2]{ForneyMonograph} for details. We omit this analysis for the sake of
simplicity.


\subsection{Sufficient condition}
In this subsection, we derive a sufficient condition on the
probability of decoding error of the code $\code_{in}$ for providing 
a positive error exponent for the code $\Code_{cont}$ as defined in subsection~\ref{sec:main-settings}.  
Below, we use the notation $\code_{in}\left[ R_{in}, n_{in} \right]$ for the code $\code_{in}$ 
of rate $R_{in}$ and length $n_{in}$. 

\begin{theorem}
\label{theorem}
Consider the BSC, and let $C$ be its capacity. Suppose that the following two conditions hold: 
\begin{itemize}
\item[(i)] There exist constants $\bet > 0$, $\vartheta > 0$, $\varepsilon_1 \in (0,1)$, 
such that for any $\epsilon$, $0 < \epsilon < \varepsilon_1$, and for a sequence 
of alphabets $\{\Phi_i\}_{i=1}^\infty$ 
where the sequence $\{ \log_2 |\Phi_i| \}_{i=1}^\infty$ is dense, there exists 
a family of codes $\Code_\Phi$ of rate $1 - \epsilon$ (with their respective decoders) that can correct a
fraction $\vartheta \epsilon^\bet$ of errors. 
\item[(ii)] There exist constants $\varepsilon_2 \in (0,1)$ and $h_0 > 0$, such that
for any $\epsilon$, $0 < \epsilon < \varepsilon_2$ , the decoding error probability of a family of codes  
$\code_{in}$ satisfies 
\[
\Prob_e\left( \code_{in} \left[ (1 - \epsilon) C, \;
\frac{1}{\epsilon^{h_0}} \right] \right)
< \epsilon^\bet \; .
\]
\end{itemize}
Then, for any rate $R < C$, there exist a family of the codes $\Code_{cont}$ as defined in 
subsection~\ref{sec:main-settings} (with respective decoder) 
that has an exponentially decaying (in $N_{cont}$) error probability.  
\end{theorem}
{\bf Proof.}
Let $R = (1 - \varepsilon) C$ be a design rate of the code $\Code_{cont}$,
and $\varepsilon > 0$ be small (namely, $\varepsilon < \min \{ \varepsilon_1, \varepsilon_2\}$). 
Let $\kappa$ be a constant, $0 < \kappa < 1$, which will be defined later, and let
the rate of the code $\code_{in}$ be $R_{in} = (1 - \kappa \, \varepsilon) C$.
We set the rate of $\Code_\Phi$ as
\[
R_\Phi = \frac{R}{R_{in}} = \frac{1 - \varepsilon}{1 - \kappa \varepsilon} 
= 1 - (1-\kappa)\varepsilon - \Theta(\varepsilon^2) \; . 
\]
Then, 
by condition (i), the fraction $\beta$ of errors correctable by the code $\Code_\Phi$ is 
at least $\beta \ge \vartheta ((1 - \kappa) \cdot \varepsilon)^\bet$. 

For an alphabet $\Phi$, the length $n_{in}$ of the code $\code_{in}$ is given by 
\begin{equation*}
n_{in} = \frac{\log_2 |\Phi|}{R_{in}} \; . 
\end{equation*}
We select the smallest $\Phi \in \{ \Phi_i \}_{i=1}^\infty$ such that 
\[
\log_2 |\Phi| \ge \frac{1}{(\kappa \varepsilon)^{h_0}} \; , 
\]
and, so, 
\begin{equation}
n_{in} > \frac{1}{(\kappa \varepsilon)^{h_0}} \; , 
\label{eq:n-in-epsilon}
\end{equation}
 

Next, we use Lemma~\ref{lemma:forney} to evaluate the decoding error probability of the code $\Code_{cont}$. 
It holds for small positive values of $\beta$ that 
\[
(1 - \beta) \ln(1 - \beta) > - \beta \; , 
\] 
and thus, from Lemma~\ref{lemma:forney} we obtain (by ignoring the positive term 
$- (1 - \beta) \ln(1 - \Prob_e (\code_{in}))$ in~(\ref{expr:forney})), 
\begin{eqnarray*}
\lefteqn{ \Prob_e \left( \Code_{cont} \right)}
\makebox[1ex]{} \nonumber \\
& < & \exp \left\{ - n \cdot \left( - \beta \ln \left( \Prob_e(\code_{in}) \right) + \beta \ln \beta
- \beta \right) \right\} \\
& = & \exp \left\{ - N_{cont} \frac{\beta}{n_{in}} \left( 
\ln \beta - \ln\left(\Prob_e(\code_{in})\right) - 1 \right) \right\} \; . 
\end{eqnarray*}

In order to have a positive error exponent, we require that 
\[
\ln \beta - \ln\left(\Prob_e(\code_{in})\right) - 1 > 0 \; , 
\]
or, equivalently, 
\begin{equation}
\beta > \e \cdot \Prob_e(\code_{in}) \; . 
\label{eq:beta-required}
\end{equation}

The decoding error probability of the selected code $\code_{in}$ satisfies: 
\begin{eqnarray}
\lefteqn{ \Prob_e \left( \code_{in} \left[ (1 - \kappa \varepsilon) C, \; n_{in} \right] \right)}
\makebox[5ex]{} \nonumber \\
& < & \Prob_e \left( \code_{in} \left[ (1 - \kappa \varepsilon) C, \; \frac{1}{(\kappa \varepsilon)^{h_0}} 
\right] \right) \nonumber \\
& < & (\kappa \varepsilon)^\bet \; 
\le \; \frac{\vartheta ((1 - \kappa) \varepsilon)^\bet}{\e} \; ,
\label{eq:prob-c-in}
\end{eqnarray}
where the first inequality is due to~(\ref{eq:n-in-epsilon}), the second inequality follows from condition (ii), 
and the third inequality can be satisfied by a selection of a small constant $\kappa$ such that
$\kappa^\bet \le \vartheta (1-\kappa)^\bet / \e$. 

The inequality~(\ref{eq:prob-c-in}) implies~(\ref{eq:beta-required}), as required. \qed


{\bf Example.} Suppose that the decoding error probability of the code $\code_{in}$ of rate 
$R_{in} = (1 - \varepsilon) C$ and length $n_{in}$ (for some decoder) is bounded by 
\[
\Prob_e(\code_{in}) < \frac{1}{n_{in}} \cdot \frac{1}{\varepsilon^4} \; . 
\]

We choose $h_0 = \bet + 5$ (where $\bet$ is as in condition (i) of Theorem~\ref{theorem}).   
There obviously exists $\varepsilon_2$ such that for every $0 < \epsilon < \varepsilon_2$,
for the code $\code_{in}$ of length $n_{in}=1/\epsilon^{h_0}$ and rate $R_{in} = (1 - \epsilon)C$,  
\begin{equation}
\Prob_e(\code_{in}) < \frac{1}{n_{in}} \cdot \frac{1}{\epsilon^4} 
= \epsilon^{h_0} \cdot \frac{1}{\epsilon^4} = \epsilon^{\bet+1} 
< \epsilon^\bet \;.
\label{eq:prob-example-1}
\end{equation}
  From the expression~(\ref{eq:prob-example-1}) we see that condition (ii) of Theorem~\ref{theorem} 
is satisfied. 
This selection guarantees existence of a positive
error exponent for the code $\Code_{cont}$. 

{\bf Example.} Suppose that the decoding error probability of the code $\code_{in}$ (of rate 
$R_{in} = (1 - \varepsilon) C$ and length $n_{in}$) is bounded by 
\[
\Prob_e(\code_{in}) < \e^{ - n_{in} \varepsilon^2 }. 
\]
We choose $h_0 = 3$. There obviously exists $\varepsilon_2$ such that for every $0 < \epsilon < \varepsilon_2$, 
for the code $\code_{in}$ of length $n_{in}=1/\epsilon^{h_0}$ and rate $R_{in} = (1 - \epsilon)C$, 
and for every $\bet>0$, 
\[
\Prob_e(\code_{in}) < \e^{ - n_{in} \epsilon^2 }
= \e^{-(\epsilon^2 /\epsilon^3)}  = \e^{-(1/\epsilon)} 
< \epsilon^\bet \; , 
\]
and therefore Theorem~\ref{theorem} yields existence of a positive error exponent
for the code $\Code_{cont}$.

\subsection{Example}

In this subsection, we consider a specific case of decoding error probability for the code $\code_{in}$. 
Theorem~\ref{theorem} can be directly applied in this case. However, we conduct a direct minimization of 
the decoding error probability of the code $\Code_{cont}$, which is obtained by concatenation of the code 
$\Code_\Phi$ in~\cite{Roth-Skachek-2004-full} with the assumed code $\code_{in}$, and obtain an analytical expression
on the error exponent. We show that the overall decoding 
error probability for this code $\Code_{cont}$ has a positive error exponent. 

Suppose that the decoding error probability for some inner code $\code_{in}$ 
over the binary symmetric channel with crossover 
probability $p < \entropy_2^{-1} (1 - R_{in})$ and some polynomial decoder is given by:
\[
\Prob_e(\code_{in}) \le \frac{1}{n_{in}^t}, 
\] 
where $t$ is a constant, $t \ge 1$. 

Below, we make a selection of parameters for the code $\Code_{cont}$. This selection 
allows us to estimate a decoding error exponent as a function of $\varepsilon$. 


Let $R = (1 - \varepsilon)C$ be a design code rate. 
Pick the rate of $\code_{in}$ to be $R_{in}=(1 - \kappa \, \varepsilon) C$,
where $\kappa \in (0,1)$ is a constant. Then, we can write 
\[
\frac{R}{R_{in}} = \frac{C(1-\varepsilon)}{C(1-\kappa \, \varepsilon)} \ge 
1 - (1 - \kappa)\varepsilon - \Theta(\varepsilon^2) \; .
\] 
Next, we select the parameters of the code $\Code_\Phi$ in~\cite{Roth-Skachek-2004-full},
which serves as an outer code. Take $\code_A$ and $\code_B$ as GRS codes over $\ff$, with $|\ff| = \Delta$. 
We fix $\delta_B = 1 - R/R_{in} - \delta_A = \eta (1 - R/R_{in})$, where $\eta \in (0,1)$ 
(and thus, $\delta_A = (1 - \eta) (1 - R/R_{in})$),  
and select the degree $\Delta$ of the graph $\graph$ as $\Delta = \varrho/\varepsilon^2$, 
where $\varrho$ is a constant, such that 
\begin{equation*}
\varrho > \frac{16}{\eta (1 - \eta) (1 - \kappa)^2}  \;  . 
\end{equation*}
We have, 
\[
R_\Phi \ge r_A + r_B - 1 = 1 - \delta_A - \delta_B = R/R_{in} \; .  
\]
By our selection (see~(\ref{eq:ramanujan})), 
\[
\gamma_\graph \le \frac{2}{\sqrt{\Delta}}
= \frac{2 \varepsilon}{\sqrt{\varrho}}\; .
\]
We obtain from~(\ref{eq:expander-distance}), 
\begin{eqnarray}
\beta > (\delta_B/2) - \gamma_\graph \sqrt{\delta_B/\delta_A} > \vartheta \varepsilon + o(\varepsilon) \; , 
\label{eq:delta}
\end{eqnarray}
where 
\[
0 < \vartheta = 
\frac{\eta (1 - \kappa)}{2} - 2 \sqrt{\frac{\eta}{\varrho(1-\eta)}} \; 
\]
is a constant which depends only on $\kappa$, $\eta$ and $\varrho$. 

The number of bits needed to represent each symbol of $\Phi$ is $\log_2 |\Phi| = r_A \Delta \cdot \log_2|\ff|$. 
Recall that $r_A = 1 - O(\varepsilon)$. Therefore, the length $n_{in}$ of the binary code $\code_{in}$ is given by
\begin{eqnarray}
n_{in} & = & \frac{r_A \Delta}{R_{in}} \cdot \log_2 (\Delta) \nonumber \\
& = & 
\frac{(1 - O(\varepsilon)) \varrho}{R_{in} \, \varepsilon^2} \; \cdot \; 
\log_2 \left( \frac{\varrho}{\varepsilon^2} \right) \nonumber \\
& = & 
\frac{\varrho \log_2(\varrho/\varepsilon^2)}{R_{in} \, \varepsilon^2} 
\;+\; o\left( \frac{\varrho \log_2(\varrho/\varepsilon^2)}{R_{in} \, \varepsilon^2} \right) \; ,
\label{eq:n-in}
\end{eqnarray}
and thus, by ignoring the small term, the decoding error probability of $\code_{in}$ is 
\begin{equation}
\Prob_e(\code_{in}) \le \left( \frac{\varepsilon^2 R_{in}}{\varrho \log_2(\varrho/\varepsilon^2)} \right)^t \; . 
\label{eq:prob-e}
\end{equation}

We substitute the expressions in~(\ref{eq:delta}) (only the main term) and~(\ref{eq:prob-e}) 
into the result of Lemma~\ref{lemma:forney} to obtain
\begin{eqnarray}
&& \Prob_e(\Code_{cont}) \;\;  < \nonumber \\
&& \qquad \exp \Bigg\{ - n \Bigg( - \vartheta \varepsilon \cdot t 
\ln \left( \frac{\varepsilon^2 R_{in}}{\varrho \log_2 (\varrho/\varepsilon^2)} \right) \nonumber \\
&& \qquad - \; (1 - \vartheta \varepsilon) \ln \left(1 - \left( \frac{\varepsilon^2 R_{in}}
{\varrho \log_2 (\varrho/\varepsilon^2)} \right)^t \right) \nonumber \\
&& \qquad + \; \vartheta \varepsilon \ln \left(\vartheta \varepsilon\right) + 
\left(1 - \vartheta \varepsilon \right) \ln \left(1 - \vartheta \varepsilon \right) \Bigg) \Bigg\} \, . 
\label{eq:exponent-forney}
\end{eqnarray}

Note that for small $\varepsilon>0$, 
\[
\ln(1 - \vartheta \varepsilon) = - \vartheta \varepsilon + O(\varepsilon^2) \; , 
\]
and 
\[
\ln \left(1 - \left( \frac{\varepsilon^2 R_{in}}{\varrho \log_2 (\varrho/\varepsilon^2)} \right)^t \right)
= - o(\varepsilon^{2t}) \; .
\]
Hence, the equation~(\ref{eq:exponent-forney}) (when neglecting $o(\varepsilon)$ terms) becomes
\begin{eqnarray*}
&& \Prob_e(\Code_{cont}) \; < \\
&& \; \; \exp \bigg\{ - n \vartheta \varepsilon  \bigg( - t 
\ln \left( \frac{\varepsilon^2 R_{in}}{\varrho \log_2 (\varrho/\varepsilon^2)} \right)  \\
&& \qquad \qquad \qquad \qquad \qquad \qquad + \ln \left(\vartheta \varepsilon\right) - 1 \bigg) \bigg\} \\
&& = \; \; \exp \left\{ - \frac{N_{cont} \vartheta \varepsilon}{n_{in}} \cdot \ln \left(\frac{\vartheta \varepsilon \cdot \varrho^t (\log_2(\varrho/\varepsilon^2))^t}{\e \cdot \varepsilon^{2t} \, R_{in}^t}\right) \right\} 
\, . 
\end{eqnarray*}
Using substitution of the expression~(\ref{eq:n-in}) for $n_{in}$, the latter equation can be rewritten as 
\begin{eqnarray}
&& \Prob_e(\Code_{cont}) \; \; < \nonumber \\
&& \quad \exp \bigg\{ - \frac{N_{cont} \vartheta \varepsilon \cdot \varepsilon^2 \, R_{in}}
{2 \varrho \, \left( \log_2(1 / \varepsilon) + \Theta(1) \right)}  \nonumber \\
&& \qquad \cdot \;\;  \bigg( (2t-1) \ln (1/\varepsilon) + t \ln(1/R_{in}) \nonumber \\
&& \qquad \qquad \qquad + \;\; t \ln \ln (1/\varepsilon) + \Theta(1) \bigg) \bigg\} \, . 
\label{eq:exponent02}
\end{eqnarray}
The dominating term in the expression 
\[
(2t-1) \ln (1/\varepsilon) + t \ln(1/R_{in}) + t \ln \ln (1/\varepsilon) + \Theta(1)
\]
is $(2t-1) \ln (1/\varepsilon)$.
By taking into account that $R_{in} = C (1 - O(\varepsilon))$, the equation~(\ref{eq:exponent02}) can be rewritten,
when ignoring all but the main term, as
\begin{eqnarray*}
&& \Prob_e(\Code_{cont}) <  \nonumber \\
&& \quad \exp \bigg\{ - N_{cont} \cdot \left( \frac{  (2t-1) \, \vartheta \, \varepsilon^3 \, C  }
{2 \varrho \cdot \log_2 \e }  + o(\varepsilon^3) \right) \bigg\} \, . \nonumber
\end{eqnarray*}
Thus, the decoding error probability is given by 
\[
\Prob_e(\Code_{cont}) < \exp \{ - N_{cont} \cdot E(C, \varepsilon) \} \; , 
\]
where 
\begin{eqnarray}
E(C, \varepsilon) & = & \max_{\varrho, \vartheta} \left\{ \frac{  \vartheta} 
{\varrho} \right\} \cdot \frac{(2t-1) \, C}{2 \cdot \log_2 \e} \cdot \varepsilon^3 \nonumber \\
& = & \max_{ \kappa , \; \eta, \; \varrho} 
\left\{ \frac{\eta (1 - \kappa)}{2 \varrho} -
2 \sqrt{\frac{\eta}{\varrho^3(1-\eta)}} \right\} \nonumber \\
&& \qquad \qquad \cdot \frac{(2t-1) \, C}{2 \cdot \log_2 \e} \cdot \varepsilon^3 \; ,
\label{eq:exponent-max}
\end{eqnarray}
and the parameters $(\kappa, \eta, \varrho)$ are taken over
\begin{equation} 
\kappa \in (0,1) \; ; \; \eta \in (0,1) \;; \; \varrho > \frac{16}{\eta (1 - \eta) (1 - \kappa)^2} \; . 
\label{eq:parameters}
\end{equation}

Next, we optimize the value of the constant
\[
\Upsilon = \max_{ \kappa , \; \eta, \; \varrho} \left\{  \frac{\eta (1 - \kappa)}{2 \varrho} -
2 \sqrt{\frac{\eta}{\varrho^3(1-\eta)}} \right\} \; . 
\]
It is easy to see that the maximum is received 
for $\kappa \rightarrow 0$. We substitute $\kappa = 0$ in expression~(\ref{eq:exponent-max}) to obtain 
\begin{eqnarray}
\Upsilon = \max_{ \eta, \; \varrho} \left\{ \frac{\eta}{2 \varrho} -
2 \sqrt{\frac{\eta}{\varrho^3(1-\eta)}} \right\} \; .
\label{eq:exponent-max-2}
\end{eqnarray}
By taking a derivative of $\Upsilon$ over $\varrho$ and comparing it to zero, we obtain that
\[
\varrho = \frac{36}{\eta(1-\eta)} \; . 
\]
By substituting it back to the expression~(\ref{eq:exponent-max-2}) and finding its maximum, we have 
$\eta = 2/3$ and $\varrho = 162$. These values obviously satisfy condition~(\ref{eq:parameters}). 
The appropriate value of $\Upsilon$ is then 
\begin{eqnarray*}
\Upsilon & = & \frac{\eta}{2 \varrho} - 2 \sqrt{\frac{\eta}{\varrho^3(1-\eta)}} =
\frac{2/3}{2 \cdot 162} - 2 \sqrt{\frac{2/3}{162^3 \cdot (1/3)}} \\
& = & \frac{1}{1458} = 6.8587 \cdot 10^{-4} \; . 
\end{eqnarray*}
Finally, we have 
\[
E(C, \varepsilon) = \frac{(2t-1) \, C}{2916 \cdot \log_2 \e} \cdot \varepsilon^3 \; . 
\]

\begin{figure}[hbt]
\begin{center}
\epsfxsize=45ex
\epsfbox{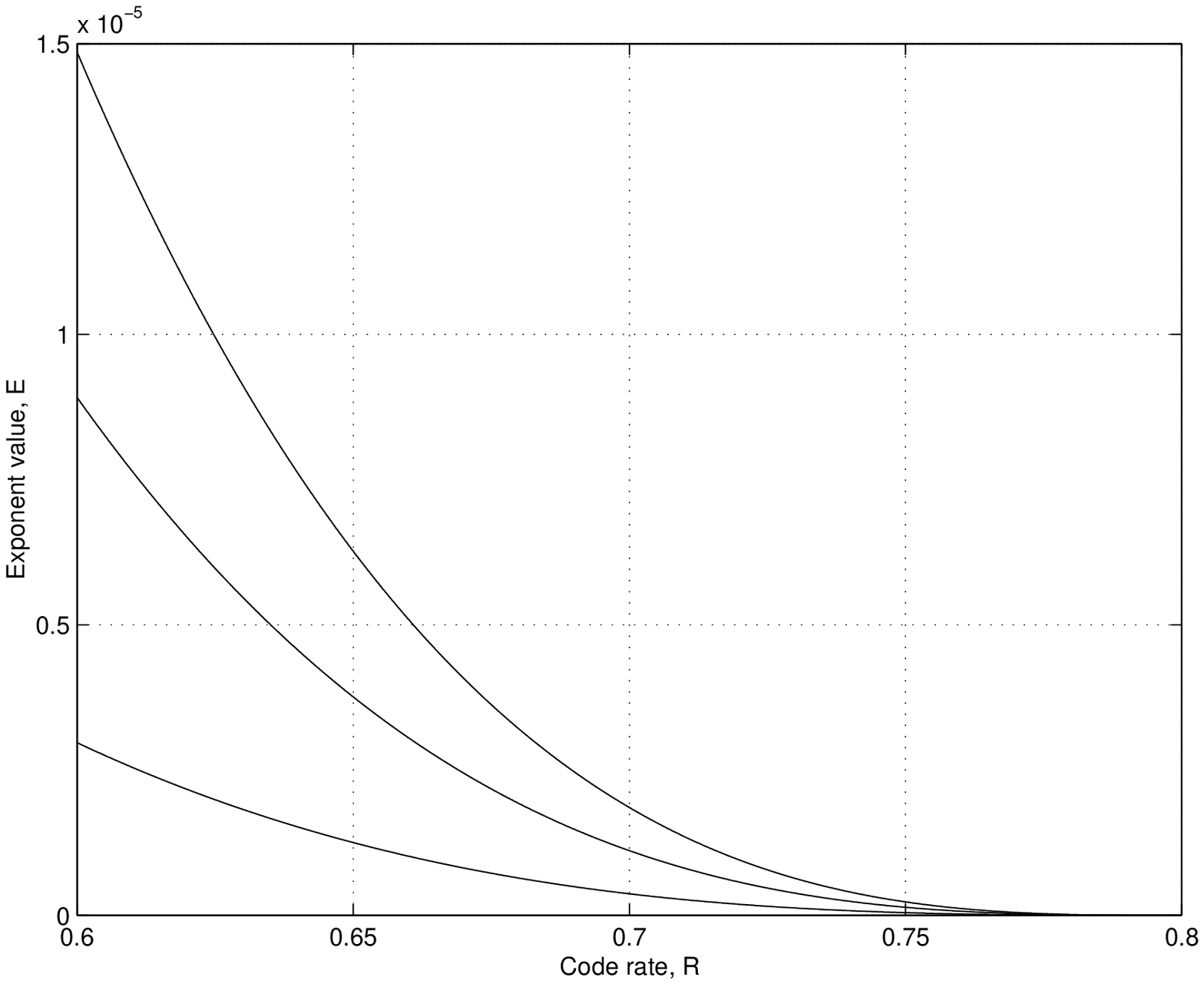}
\end{center}
\caption{Error exponent $E(C,\varepsilon)$ for the code $\Code_{cont}$.}
\vspace{-1.5ex}
\begin{center}
{\footnotesize Selection: $\Prob_e(\code_{in}) = 1/n_{in}^t$; $\; C=0.8$; $\; t=1,2,3$ (bottom to top).}
\end{center}

\label{fig:exponents}
\end{figure}

Figure~\ref{fig:exponents} shows value of error exponent $E(C, \varepsilon)$ in the example for $t=1$, $2$ and $3$. 

\subsection{Decoding complexity}
In this subsection, we show that under the assumption in Section~\ref{sec:assumed-code} on the
decoding time complexity of the code $\code_{in}$, and if the parameters of the codes
are selected as in the proof of Theorem~\ref{theorem}, then the decoding time complexity
of the respective code $\Code_{cont}$ is linear in the overall length $N_{cont}$ and 
inverse polynomial in the gap from capacity $\varepsilon$.  

\begin{theorem}
\label{theorem:complexity}
Consider the BSC, and let $C$ be its capacity. Let $R = (1 - \varepsilon) C$ be a design rate. 
Suppose that the following two conditions hold: 
\begin{itemize}
\item[(i)] Let $\Code_\Phi$ be a (family of) code defined in Section~\ref{sec:expander-code}
of rate $R_\Phi = (1 - \varepsilon)/(1 - \kappa \varepsilon)$, $\kappa \in (0,1)$ is a constant, 
over a smallest alphabet $\Phi$ satisfying $\log_2 |\Phi| \ge 1/(\kappa \varepsilon)^{h_0}$ 
from a dense sequence $\{ \log_2 |\Phi_i| \}_{i=1}^\infty$, 
and $h_0>0$ is a constant. 

\item[(ii)] Let $\code_{in}$ be a code of rate $R_{in} = (1 - \kappa \varepsilon)C$ with 
the decoding complexity over the BSC of capacity $C$ given by 
\begin{equation*}
O\left(n_{in}^{\ssf} \cdot \frac{1}{\varepsilon^\rsf}\right) \; ,
\end{equation*}
where $\ssf, \rsf \ge 1$ are some constants. 
\end{itemize}
Then, the time complexity of the respective code $\Code_{cont}$, when decoded 
by $\decoder_{cont}$, is given by 
\[
N_{cont} \cdot \mbox{\sc Poly}(1/\varepsilon) \; . 
\] 
$\,$
\end{theorem}

{\bf Proof.} 
Below we count the total number of operations when decoding the code $\Code_{cont}$
by the decoder $\decoder_{cont}$.
There are two main steps.   
\begin{itemize}
\item{Step 1:} $n$ applications of the decoder $\decoder_{in}$ 
on the binary word of length $n_{in}$.
\item{Step 2:} one application of the decoder $\decoder_\Phi$ on the word of length $n$ over $\Phi$. 
\end{itemize}
In addition, there are $n$ applications of each of the mappings $\encoder_0$ and $\encoder_0^{-1}$. 

We separately count the number of operations during each step. 
\begin{itemize}
\item{Step 1:} By the assumption on the decoding complexity of $\decoder_{in}$, 
$n$ applications of this decoder result in time
\begin{equation}
O \left(n \cdot n_{in}^{\ssf} \cdot \frac{1}{\varepsilon^\rsf}\right) =
O \left(N_{cont} \cdot n_{in}^{\ssf-1} \cdot \frac{1}{\varepsilon^\rsf}\right) \; .
\label{eq:complexity-step1}
\end{equation}

  From the definition of $\Code_{cont}$, $n_{in} = \log_2|\Phi| \,/\, R_{in}$, so, we have
\[
n_{in} = \frac{\log_2 |\Phi|}{(1 - \kappa \varepsilon)C} \; .
\]
By using the density of values of $\log_2 |\Phi|$, we have $\log_2 |\Phi| \in \mbox{\sc Poly}(1/\varepsilon)$, 
thus yielding 
$n_{in} \in \mbox{\sc Poly}(1/\varepsilon)$.  
By substitution into~(\ref{eq:complexity-step1}), 
we obtain that the
time complexity of Step 1 is $N_{cont} \cdot \mbox{\sc Poly}(1/\varepsilon)$.
\item{Step 2:} it is shown in~\cite{Roth-Skachek-2004-full} that the number of applications of decoders
$\decoder_A$ and $\decoder_B$ on the word of $\Code_\Phi$ 
of length $n$ over $\Phi$ is bounded by $\omega \cdot n$, where
\[
\omega = 2 \cdot \left\lceil
\frac{\ln\left( \frac{\displaystyle
\Delta \beta \sqrt{\sigma}}{ \displaystyle {\beta} - \sigma} \right)}
{\ln \left( \displaystyle
\frac{\delta_A \delta_B}{4 \gamma_\graph^2} \right) } \right\rceil
+ \frac{\displaystyle 1 + \frac{\delta_A}{\delta_B} } 
{\displaystyle 1 -
\left( \frac{4 \gamma_\graph^2}{\delta_A \delta_B} \right)^2 } \; ,
\]
and $\sigma$ is an actual number of errors in the word. Thus, if the ratio $\sigma/\beta$ is bounded away 
from $1$, and $\graph$ is a Ramanujan graph, then  the value of $\omega$ is bounded from above by 
an absolute constant (independent of $\Delta$).  

The decoders $\decoder_A$ and $\decoder_B$ are applied on the words of length 
$\Delta \in \mbox{\sc Poly}(1/\varepsilon)$. When half minimum distance decoders
for GRS codes are used, their complexity is polynomial in $1/\varepsilon$. Therefore, 
the decoding complexity in Step 2 is bounded by 
\[
n \cdot \mbox{\sc Poly}(1/\varepsilon) \le N_{cont} \cdot \mbox{\sc Poly}(1/\varepsilon) \; . 
\]
\end{itemize}

Each application of mapping $\encoder_0$ or $\encoder_0^{-1}$ is equivalent 
to multiplication of a vector by a matrix, where the number of rows and 
columns in the matrix is $\mbox{\sc Poly}(1/\varepsilon)$. This can be done in 
time $\mbox{\sc Poly}(1/\varepsilon)$.  

Summing up the decoding complexities of all steps of the decoder, we obtain that the 
total number of operations is bounded by 
\[
N_{cont} \cdot \mbox{\sc Poly}(1/\varepsilon) \; . 
\]
\qed

{\bf Note.} The result in Theorem~\ref{theorem:complexity} is still valid if the outer code 
$\Code_\Phi$ be replaced by any other code of rate $1 - \Theta(\varepsilon)$, 
whose decoding time complexity is linear in $n$ and polynomial in $1/\varepsilon$, 
for a $\log$-dense sequence of alphabet sizes. 

\section{Time complexity of decoder in~\cite{Zemor02}} 
\label{sec:analysis-zemor02}

Similarly to Section~\ref{sec:main-results}, assume in this and the next sections that $C$ is the capacity 
of the BSC with crossover probability $p$, and the design code rate is $R=(1-\varepsilon)C$.
Our purpose is to compare the parameters of the codes from Section~\ref{sec:main-results} 
with codes presented by Barg and Z\'emor in~\cite{Zemor02} and~\cite{Zemor03} (with their respective decoding algorithms). In the sequel we show 
that the parameters of the codes from~\cite{Zemor02} and~\cite{Zemor03}
cannot be modified such that the decoding time complexity would be only 
sub-exponential in $1/\varepsilon$ while keeping a non-zero error exponent. The reason is this: both decoding 
algorithms in~\cite{Zemor02} and~\cite{Zemor03} make use of sub-routines (decoders for small constituent 
codes) that have time complexity exponential in a degree of underlying expander graph. This degree, in turn, 
depends (at least) polynomially on $1/\varepsilon$. 
 
\subsection{Construction}

We briefly recall the construction and the decoder in~\cite{Zemor02}.  
Let $\graph = (A:B,E)$ be 
a bipartite $\Delta$-regular undirected connected graph with
a vertex set $V = A \cup B$ such that
$A \cap B = \emptyset$ and $|A| = |B| = n$,
and an edge set $E$ of size $N = \Delta n$ such that
every edge in $E$ has one endpoint in $A$ and one endpoint in $B$. 

Let the size of the finite field $\ff $ be a power of $2$. 
Let $\code_A$ and $\code_B$ be two \emph{random} codes of length $\Delta$ over $\ff$.   
The code $\Code_{BZ2} = (\graph, \code_A : \code_B)$ is defined similarly to the definition of $\Code$ 
in~(\ref{eq:define_c}), with respect to $\code_A$ and $\code_B$ as defined in this paragraph.

\subsection{Decoding}

Let us submit a word $\bldc = (c_e)_{e \in E} \in \Code_{BZ2}$ to the BSC. Assume that $\bldy = (y_e)_{e \in E}$
is a received (erroneous) word. A formal definition of the decoder $\decoder_{BZ2}$ appears in Figure~\ref{fig:decoder-BZ2}.
\begin{figure}[hbt]
\makebox[0in]{}\hrulefill\makebox[0in]{}
{\normalsize
\begin{description}
\item[\underline{\bfseries Input:}]
$\,$ Received word $\bldy = (y_e)_{e \in E}$
in $\ff^N$.
\item[\underline{\bfseries Let}] $\bldz \leftarrow \bldy$. 
\item[\underline{\bfseries For}] $i \leftarrow 1,2,\ldots,m$
\underline{\bfseries do} \{
\begin{description}
\item[\underline{\bfseries If}]
$i$ is odd $\;$ \underline{\bfseries then} $X \equiv A$,
$\decoder \equiv \decoder_A$, \\
\underline{\bfseries else} $X \equiv B$, $\decoder \equiv \decoder_B$.
\item[\underline{\bfseries For}]
$u \in X$ \underline{\bfseries do}
$(\bldz)_{\!\scriptscriptstyle E(u)}
\leftarrow \decoder((\bldz)_{\!\scriptscriptstyle E(u)})$.
\end{description} 
\item \}
\item[\underline{\bfseries Output:}]
$\;\;\;\; \bldz$ if $\bldz \in \Code_{BZ2}$ (and declare `error' otherwise).
\end{description}
}
\makebox[0in]{}\hrulefill\makebox[0in]{}
\caption{Decoder $\decoder_{BZ2}$ of Barg and Z\'emor for the code $\Code_{BZ2}$.}
\label{fig:decoder-BZ2}
\end{figure}
The number of iterations $m$ is taken to be $O(\log n)$. The decoders $\decoder_A$ and $\decoder_B$ are 
the \emph{maximum-likelihood} decoders for the codes $\code_A$ and $\code_B$, respectively.  

The analysis of codes in~\cite{Zemor02} is divided into two cases. In the first case, 
the codes $\code_A$ and $\code_B$ over $\ff = \GF(2)$ are considered. In the second case, 
the analysis is generalized toward field sizes, which are large powers of 2. 
We analyze these two cases separately. 

\subsection{Analysis: binary codes}

In the binary case, following the analysis of~\cite{Zemor02} 
it is possible to show that for the code $\Code_{BZ2}$ with the decoder $\decoder_{BZ2}$,
the decoding error probability, $\Prob_e(\Code_{BZ2})$, is bounded by 
\[
\Prob_e(\Code_{BZ2}, p) \le \exp\{ - \alpha N f_3(R,p) \} \; ,
\]
where $0 < \alpha < 1$, and the main term of $f_3 (R,p)$ is less or equal to
\begin{eqnarray}
\max_{R \le R_0 < C} & \left\{ E_0(R_0, p) 
\left( \frac{\entropy_2^{-1}(R_0 - R)}{2} - \Theta\left( \frac{1}{\sqrt{\Delta}} \right) \right) \right\} , 
\label{eq:zemor-2-exp}
\end{eqnarray}
and $E_0(R_0, p)$ is the \emph{random coding exponent} for rate $R_0$ over 
the BSC with a crossover probability $p$.   

\begin{proposition}
If the codes $\Code_{BZ2}$ (binary, as assumed in this subsection), 
have a positive error exponent under the decoding by $\decoder_{BZ2}$, 
then $\Delta = \Omega \left(1/(\entropy_2^{-1}(\varepsilon))^2 \right)$. 
\label{prop:exp-zemor-2}
\end{proposition}
{\bf Proof.} 
In order to have a positive error exponent it is needed that 
\[
\frac{\entropy_2^{-1}(R_0 - R)}{2} - \Theta\left( \frac{1}{\sqrt{\Delta}} \right) > 0 \; . 
\]

Observe that $R_0-R \le C-R = C \varepsilon \le \varepsilon$. 
It follows from~(\ref{eq:zemor-2-exp}) that 
\[
\half \entropy_2^{-1}(\varepsilon) \ge \half \entropy_2^{-1}(R_0 - R)  >
\Theta \left( 1/\sqrt{\Delta} \right), 
\] 
and thus $\Delta = \Omega \left( 1/(\entropy_2^{-1}(\varepsilon))^2 \right)$. $\qed$

It is suggested in~\cite{Zemor02} to use the maximum-likelihood deco\-ding for \emph{random} codes $\code_A$ and $\code_B$. This decoding, how\-ever, has time complexity at least
\[
\exp\{ \Omega(\Delta) \}  = \exp\{ \Omega\left(1/(\entropy_2^{-1}(\varepsilon))^2 \right) \} \; . 
\]

\subsection{Analysis: codes over large fields}

Suppose that the size of the field $\ff$ is a large power of 2. 
In this case, for the code $\Code_{BZ2}$ under the decoding by $\decoder_{BZ2}$, 
the decoding error probability $\Prob_e(\Code_{BZ2})$ is bounded by 
\[
\Prob_e(\Code_{BZ2}, p) \le \exp\{ - \alpha N f_2(R,p) \} \; ,
\]
and the main term of $f_2 (R,p)$ is less or equal to
\begin{eqnarray*}
\max_{R \le R_0 < C} & \left\{ E_0(R_0, p) 
\left( \frac{R_0 - R}{2} - \Theta\left( \frac{1}{\sqrt{\Delta}} \right) \right) \right\} \; . 
\end{eqnarray*}
In this case, Proposition~\ref{prop:exp-zemor-2} can be rewritten as 
\begin{proposition}
If the codes $\Code_{BZ2}$ (over large $\ff$, as assumed in this subsection)
have a positive error exponent under the decoding by $\decoder_{BZ2}$, 
then $\Delta = \Omega \left(1/\varepsilon^2 \right)$. 
\end{proposition}

The proof is very similar to that of Proposition~\ref{prop:exp-zemor-2}. 

When using the maximum-likelihood decoder for \emph{random} codes $\code_A$ and $\code_B$, 
the decoding time complexity is at least
\[
\exp\{ \Omega(\Delta) \}  = \exp\{ \Omega\left(1/\varepsilon^2 \right) \} \; . 
\]

\section{Time complexity of decoder in~\cite{Zemor03}} 
\label{sec:analysis-zemor03}
\subsection{Construction}
Recall the construction of expander codes presented in~\cite{Zemor03}. Let $\graph=(V,E)$ be a 
bipartite graph with $V = V_0 \cup (V_1 \cup V_2)$, such that each edge has one endpoint in $V_0$ and 
one endpoint in either $V_1$ or $V_2$. Let $|V_i| = n$ for $i = 0,1,2$. Let the degree of each vertex in $V_0$,
$V_1$, and $V_2$ be $\Delta$, $\Delta_1$, and $\Delta_2=\Delta-\Delta_1$, respectively. In addition, let the subgraph 
$\graph_1$ induced by $V_0 \cup V_1$ be a regular bipartite Ramanujan graph and denote by $E_1$ its edge set.   
Let $\lambda_1$ be a second largest eigenvalue of the adjacency matrix of $\graph_1$. 

Let $\code_A$ be a $[l \Delta, R_0 l \Delta, d_0 = l \Delta \delta_0]$ linear binary code of rate 
$R_0 = \Delta_1 / \Delta$. Let $\code_B$ be $q$-ary $[\Delta_1, R_1 \Delta_1, d_1 = \Delta_1 \delta_1]$ additive code,
and let $q = 2^l$. Let $\code_{aux}$ 
be $q$-ary code of length $\Delta_1$. The code $\Code_{BZ3}$ is defined as the set of vectors
$\bldx = \{ x_1, x_2, \cdots, x_N \}$, indexed by the set $E$ of size $N = \Delta n$, such that  
\begin{enumerate}
\item 
For every vertex $v \in V_0$, the subvector $(x_j)_{j \in E(v)}$ is a $q$-ary codeword of $\code_A$ and the set
of coordinates $E_1(v)$ is an information set for the code $\code_A$. 
\item
For every vertex $v \in V_1$, the subvector $(x_j)_{j \in E(v)}$ is a $q$-ary codeword of $\code_B$.
\item
For every vertex $v \in V_0$, the subvector $(x_j)_{j \in E_1(v)}$ is a codeword of $\code_{aux}$.
\end{enumerate}

\subsection{Decoding}
The authors of~\cite{Zemor03} proposed decoding algorithm for the code $\Code_{BZ3}$. 
In the first iteration, each
subvector $\bldz(v)$, $v \in V_0$, is treated as following: the decoder computes, for every symbol
$b$ of the $q$-ary alphabet, and for every edge $e \in E_1$ incident to $v$, the weight of the edge as
follows:
\[
d_{e,b}(\bldz) = \min_{\blda \in \code_A: a_e = b} \dist(\blda, \bldz(v)),
\]  
where $a_e$ denotes the $q$-ary coordinate of the codeword $\blda$ that corresponds to the edge $e$,
and $\dist(\cdot, \cdot)$ is the binary Hamming distance. This information is passed along the edge $e$ 
to the corresponding decoder on the right-hand side of the bipartite graph. In the second iteration, 
for every vertex $w \in V_1$ the right decoder associated to it finds a $q$-ary codeword 
$\bldb = (b_1, \ldots, b_{\Delta_1}) \in \code_B$ that satisfies
\[
\bldb = \arg \min_{\bldx = (x_1, \ldots, x_{\Delta_1}) \in \code_B}
 \sum_{i=1}^{\Delta_1} d_{w(i), x_i}(\bldz) \;  ,
\]
and writes $b_i$ on the edge $w(i)$, $i = 1, \ldots, \Delta_1$. 

Then, the decoder continues similarly to the decoder in~\cite{Zemor02}. 

\subsection{Analysis} 
\begin{lemma}
Let $p$ satisfy $0 < p < \half$, and let $0 < \varepsilon \ll p$. Then, 
\begin{eqnarray*}
\entropy_2^{-1} \left( \entropy_2(p) + \varepsilon ( 1 - \entropy_2(p)) \right) = 
p + \frac{ \varepsilon ( 1 - \entropy_2(p))}{\log_2 \left( (1-p)/p \right)} \\ - 
\frac{ \varepsilon^2 ( 1 - \entropy_2(p))^2 \log_2 \e}{2 p(p-1) \left( \log_2 \left( (1-p)/p \right) \right)^3}  + O(\varepsilon^3).
\end{eqnarray*}
\label{lemma:entropy-1}
\end{lemma}
The proof of this lemma appears in the Appendix B.
  
\begin{proposition}
Let $C$ be the capacity of the BSC.
The decoding error probability of  
a random code of rate $R=(1-\varepsilon )C$, under the maximum-likelihood decoding, 
behaves as $\exp\{ - \Theta (\varepsilon^2) \}$ when $\varepsilon
\rightarrow 0$. 
\label{prop:epsilon-2} 
\end{proposition}
{\bf Proof.}
We start with the well-known expression for the probability exponent of the decoding error of a random 
code under the maximum-likelihood decoding~\cite{Gallager-monograph},~\cite{Gallager}.
\begin{eqnarray*}
&& \hspace{-5ex} E_0(R, p) = \nonumber \\
&& \hspace{-5ex} \left\{ \begin{array}{ll}
T(\delta, p) + R  - 1 & \hspace{-1ex} \mbox{ if } R_{crit} \le R < C \\
1 - \log_2 \left( 1 + \sqrt{4p(1-p)} \right) - R & \hspace{-1ex} \mbox{ if } R_{min} \le R < R_{crit} \\
- \delta \log_2 \sqrt{4p(1-p)} & \hspace{-1ex} \mbox{ if } 0 \le R < R_{min} \; , 
\end{array} \right. 
\label{eq:decod_error} 
\end{eqnarray*}
where $R_{min}$ and $R_{crit}$ are some threshold rates, 
\[
\delta = \delta_{GV}(R) = \entropy_2^{-1} (1 - R) \; , 
\]
and 
\[
T(x,y) = - x \log_2 y - (1 - x) \log_2 (1 - y) \; . 
\]
At the code rates $R$ which are close to $C$, the relevant expression for random coding exponent
becomes 
\begin{equation}
E_0(R, p) = T(\delta, p) + R  - 1 \; .
\label{eq:exp-random}
\end{equation}

Next, we express all terms of the relevant part of~(\ref{eq:exp-random}) in terms of $\varepsilon$. We recall, that 
$R = (1 - \varepsilon) ( 1 - \entropy_2(p))$ and, thus, 
\[
\entropy_2^{-1} (1 - R) = \entropy_2^{-1} (\varepsilon + \entropy_2(p) - \varepsilon \entropy_2(p)) \; .
\] 
Thus, when disregarding $O(\varepsilon^3)$ term, the equation~(\ref{eq:decod_error}) becomes

\begin{eqnarray}
&& \hspace{-6ex} E_0(R,p) = \nonumber \\
&& (1 - \varepsilon) ( 1 - \entropy_2(p)) - 1 \nonumber \\
&& \quad + T \left(\entropy_2^{-1} (\varepsilon + \entropy_2(p) - \varepsilon \entropy_2(p)), \; p \right) \nonumber \\
& \stackrel{(*)}{=} & - \varepsilon - (1 - \varepsilon) \entropy_2(p) + T \Bigg(p + \frac{\varepsilon ( 1 - \entropy_2(p))}{\log_2((1-p)/p)} \nonumber \\
&& \quad - \frac{\varepsilon^2 (1 - \entropy_2(p))^2 \log_2 \e}{2 p(p-1)\left(\log_2((1-p)/p)\right)^3}, 
\; p \Bigg) \nonumber
\end{eqnarray}
\vspace{-2ex}
\begin{eqnarray}
& = & - \varepsilon - (1 - \varepsilon) \entropy_2(p) \nonumber - \Bigg(p + \frac{\varepsilon ( 1 -
\entropy_2(p))}{\log_2((1-p)/p)} \nonumber \\
&& \quad - \frac{\varepsilon^2 (1 - \entropy_2(p))^2 \log_2 \e}{2 p(p-1)\left(\log_2((1-p)/p)\right)^3} \Bigg) \log_2 p \nonumber \\
&  & \quad - \Bigg( 1 - p - \frac{\varepsilon ( 1 - \entropy_2(p))}{\log_2((1-p)/p)} \nonumber \\
&& \quad + \frac{\varepsilon^2 (1 - \entropy_2(p))^2 \log_2 \e}{2 p(p-1)\left(\log_2((1-p)/p)\right)^3}\Bigg) 
\log_2(1 - p ) \nonumber  
\end{eqnarray}
\vspace{-2ex}
\begin{eqnarray*}
& = & - \varepsilon (1 - \entropy_2(p))  \nonumber \\
&& \quad + \frac{\varepsilon (1 - \entropy_2(p))(- \log_2 p + \log_2 (1-p) )}{\log_2((1-p)/p)} 
 \nonumber \\
&& \quad + \frac{\varepsilon^2 (1 - \entropy_2(p))^2 \log_2 \e ( \log_2 p - \log_2 (1-p))}
{2 p(p-1)\left(\log_2((1-p)/p)\right)^3}
 \nonumber \\
& = & \frac{\varepsilon^2 (1 - \entropy_2(p))^2 \log_2 \e}{2 p(1-p)\left(\log_2((1-p)/p)\right)^2} = 
\varepsilon^2 \cdot c_p \; , 
\end{eqnarray*} 
where $c_p>0$ is a constant that depends only on the crossover probability $p$ of the channel. Note that the transition
$(*)$ follows from Lemma~\ref{lemma:entropy-1}. $\qed$

\begin{proposition}
If the codes $\Code_{BZ3}$ have a positive error exponent, then $\Delta = \Omega(1/\varepsilon^2)$.
\end{proposition}
{\bf Proof.}
It is shown in~\cite{Zemor03} that the decoding error probability of the code $\Code_{BZ3}$, 
$\Prob_e(\Code_{BZ3})$, satisfies
\begin{eqnarray*}
\Prob_e(\Code_{BZ3}) \le \exp \left\{
-n \Delta l \delta_1 (1 + \alpha)^{-1} \right. \\
\left. \cdot (E_0(R_0,p) - M \alpha) (1 - o(1)) \right\},  
\end{eqnarray*}
where $\alpha$ is a constant defined in~\cite{Zemor03} (in paritcular, $1 > \alpha > 2\lambda_1/d_1$), and 
\[
M = M(R,p) = \left\{ \begin{array}{ll}
\frac{1}{2} \log_2 ((1-p)/p) & \mbox{ if } R \le R_{crit} \\
\log_2 \left( \frac{\delta_{GV}(R) (1-p)}{(1 - \delta_{GV}(R))p} \right) & \mbox{ if } R \ge R_{crit}
\end{array} \right. \; , 
\] 
$\delta_{GV}(R) = \entropy_2^{-1} (1-R)$ is the Gilbert-Varshamov relative distance for the rate R, and
$R_{crit} = 1 - \entropy_2(\rho_0)$ is a so-called \emph{critical rate}, where $\rho_0 = \sqrt{p}/(\sqrt{p}+\sqrt{1-p})$ (see~\cite{Zemor03} for details).
 
We are interested in small values of $\varepsilon$, i.e. $R \ge R_{crit}$. In this case, the value of
$M(R,p)$ can be rewritten as 
\begin{eqnarray}
\lefteqn{M(R,p) 
\;\; = \;\; \log_2 \left( \frac{\delta_{GV}(R) (1-p)}{(1 - \delta_{GV}(R))p} \right)} \makebox[0ex]{} \nonumber \\ 
& = & \log_2 \left( \frac{\entropy_2^{-1}(1-R) (1-p)}{(1 - \entropy_2^{-1}(1-R))p} \right) \nonumber \\
& = & \log_2 \left( \frac{\entropy_2^{-1}(\entropy_2(p) + \varepsilon - \varepsilon \entropy_2(p)) (1-p)}
{(1 - \entropy_2^{-1}(\entropy_2(p) + \varepsilon - \varepsilon \entropy_2(p)))p} \right) \; ,
\label{eq:M_R_p}
\end{eqnarray}
where the last transition is due to $R = (1 - \entropy_2(p)) (1 - \varepsilon)$. 
Using Lemma~\ref{lemma:entropy-1}, the equality~(\ref{eq:M_R_p}) becomes 
\begin{eqnarray*}
&& \hspace{-5ex} M(R,p) = \\ 
&& \hspace{-4ex} \log_2 
\frac{\left( p + \frac{ \varepsilon ( 1 - \entropy_2(p))}{\log_2 \left( (1-p)/p \right)} - \frac{1}{2} \cdot
\frac{ \varepsilon^2 ( 1 - \entropy_2(p))^2 \log_2 \e}{p(p-1) \left( \log_2 \left( (1-p)/p \right) \right)^3 } \right) 
\left( 1-p \right) }
{\left( 1 - p - \frac{\varepsilon ( 1 - \entropy_2(p))}{ \log_2 \left( (1-p)/p \right)} + \frac{1}{2} \cdot
\frac{ \varepsilon^2 ( 1 - \entropy_2(p))^2 \log_2 \e}
{p(p-1) \left( \log_2 \left( (1-p)/p \right)) \right)^3 } \right) p } \\
&& \hspace{40ex} + O(\varepsilon^3)\; .
\end{eqnarray*}
When ignoring the terms of $\varepsilon^2$ and highest powers of $\varepsilon$, 
and denoting $\theta = \frac{ \varepsilon ( 1 - \entropy_2(p))}{\log_2 \left( (1-p)/p \right)}$, 
this equation becomes
\begin{eqnarray*}
M(R,p) & = & \log_2 \left( \frac{p + \theta}{1 - p - \theta} \cdot \frac{1-p}{p} \right)  + O(\theta^2) \\
& = &
\log_2 \left( \frac{1 + \theta / p }{1 - \theta/(1-p)} \right)  + O(\theta^2) \\
& = & \log_2 \left( (1 + \theta / p )(1 + \theta/(1-p)) \right) + O(\theta^2) \\
& = & \log_2 \left( 1 + \theta / p  + \theta/(1-p) \right) + O(\theta^2) \; . 
\end{eqnarray*}
Using Taylor's series for $\ln(\cdot)$ around $1$ we obtain 
\begin{eqnarray*}
M(R,p) & = & \log_2 \e \cdot
\left( \frac{\theta}{p} + \frac{\theta}{(1-p)} \right) + O(\theta^2) \\ 
& = & \frac{\log_2 \e}{p(1-p)} \cdot \theta + O(\theta^2) \; , 
\end{eqnarray*}
and switching back to $\varepsilon$ notation this becomes 
\begin{equation}
M(R,p) = \frac{\log_2 \e}{p(1-p)} \cdot \frac{ \varepsilon ( 1 - \entropy_2(p))}
{\log_2 \left( (1-p)/p \right)} + O(\varepsilon^2)= \Theta(\varepsilon) \; . 
\label{eq:M}
\end{equation}
Next, we evaluate the value of $\alpha$. Recall that $\alpha > 2\lambda_1 / d_1$, and 
$d_1 \le \Delta_1 \le \Delta$. We have 
\[
\alpha > \frac{2 \lambda_1}{d_1} \ge \frac{4 \sqrt{\Delta_1 - 1}}{\Delta_1} \ge \frac{4  \sqrt{\Delta - 1}}{\Delta} 
= \Theta \left( \frac{1}{\sqrt{\Delta}} \right) \; . 
\] 
In order to have a positive error exponent it is necessary that 
\begin{eqnarray*}
 E_0(R_0, p) - M \alpha > 0 
& \Rightarrow & \frac{E_0(R_0, p)}{M} > \alpha \\ 
& \Rightarrow & \frac{E_0(R_0, p)}{M} > \Theta \left( \frac{1}{\sqrt{\Delta}} \right)  \; . 
\end{eqnarray*}
Using Proposition~\ref{prop:epsilon-2}, $E_0(R_0, p) = \Theta(\varepsilon^2)$, and thus from~(\ref{eq:M})
\[
\varepsilon = \Omega(1/\sqrt{\Delta}) \qquad \Rightarrow  \qquad \Delta = \Omega(1/\varepsilon^2) \; .
\]
$\qed$

Assuming that the first two decoding iterations are as suggested in \cite{Zemor03}, 
we conclude that the time complexity of
the decoding is $\exp\{\Omega(\Delta)\} = \exp\{\Omega(1 / \varepsilon^2)\}$.

\section*{Appendix A} 
{\bf Proof of Lemma~\ref{lemma:forney}.} 

We analyze the error exponent, following the guidelines of the analysis of Forney~\cite[Chapter 4.2]{ForneyMonograph}. Let $\varsigma_i$, $i=1,\cdots, n$, be a random variable which equals $1$ if no inner decoding error is made while decoding $i$-th inner codeword, and $-1$ otherwise. The outer code will fail to decode correctly
if and only if 
\[
\varsigma \stackrel{\triangle}{=} \frac{1}{n} \sum_{i=1}^n \varsigma_i < (1 - 2 \beta) \; . 
\]
Denote 
\[
\mu(-s) \stackrel{\triangle}{=} 
\ln \left( \Prob_e(\code_{in}) \cdot \e^s + (1 - \Prob_e(\code_{in})) \cdot \e^{-s} \right) \; . 
\]
Using the Chernoff bound, we obtain
\begin{eqnarray*}
\Prob_e(\Code_\Phi) & = & \Prob\left( \frac{1}{n} \sum_{i=1}^n \varsigma_i <(1 - 2 \beta) \right) \\
& < & \e^{-n \left( s(2 \beta-1) - \mu(-s) \right)} \; . 
\end{eqnarray*}
Optimization of the exponent over values of $s$ yields that the maximum of the expression 
\[
s(2 \beta - 1) - \mu(-s)
\]
is achieved when 
\[
s = \half \ln \frac{(1-\Prob_e(\code_{in})) \cdot 2 \beta}{\Prob_e(\code_{in}) \cdot (2-2 \beta)} \; , 
\]
and the maximum is 
\begin{eqnarray*}
s(2 \beta-1) - \mu(-s) & = & - \; \beta \ln \left(\Prob_e(\code_{in})\right) \\
& - & \left(1 - \beta \right) \ln \left(1 - \Prob_e(\code_{in})\right) \\
& + & \beta \ln \left( \beta \right) + 
\left(1 - \beta \right) \ln \left(1 - \beta \right) \; ,
\end{eqnarray*}
thus completing the proof.  $\qed$

\section*{Appendix B} 
{\bf Proof of Lemma~\ref{lemma:entropy-1}.} 

Consider the value of the binary entropy function at the point $p+x$ for small $x>0$. 
Using Taylor series around point $p$, 
\[
\entropy_2(p + x) = \entropy_2(p) + \entropy_2'(p) \cdot x + \frac{1}{2} \entropy_2''(p) \cdot x^2 + O(x^3) \; . 
\]
By calculation of the derivatives of the entropy function, one obtains
\begin{eqnarray*}
\entropy_2'(\chi) & = & - \log_2 \chi - \chi \cdot \frac{1}{\chi} \cdot \log_2 \e + \log_2 (1 - \chi) \\
& + & (1 - \chi) \cdot \frac{1}{1-\chi} \cdot \log_2 \e = \log_2 \left( \frac{1- \chi}{\chi} \right) \; ;
\end{eqnarray*}
and 
\begin{eqnarray*}
\entropy_2''(\chi) = \log_2 \e \cdot \left( - \frac{1}{1-\chi} - \frac{1}{\chi} \right) 
= \frac{\log_2 \e}{\chi (\chi - 1)} \; .
\end{eqnarray*}
Therefore,
\begin{eqnarray*}
&& \entropy_2(p + x) = \\
&& \quad \entropy_2(p) + \log_2 \left( \frac{1-p}{p} \right) \cdot x + \frac{\log_2 \e}{p(p-1)} \frac{x^2}{2} + O(x^3) \; .
\end{eqnarray*}
By applying the inverse of the binary entropy function on both sides of the equation,  
\begin{eqnarray*}
p + x & = & \entropy_2^{-1} \left( \entropy_2(p + x) \right)\nonumber \\
& = & \entropy_2^{-1} \bigg( \entropy_2(p) + \log_2 \left( \frac{1-p}{p} \right) \cdot x \nonumber \\
&& \qquad + \frac{\log_2 \e}{p(p-1)} \cdot \frac{x^2}{2} + O(x^3) \bigg) \; . 
\end{eqnarray*}
Denote by $\theta$ the value of $\log_2 \left( \frac{1-p}{p} \right) \cdot x + \frac{\log_2 \e}{p(p-1)} \cdot \frac{x^2}{2}$, thus obtaining 
\begin{eqnarray}
p + x = \entropy_2^{-1} \left( \entropy_2(p) + \theta + O(x^3) \right) \; .
\label{eq:entropy-taylor-1}
\end{eqnarray}
By solving the quadratic equation 
\[
\theta = \left( \ln \left( \frac{1-p}{p} \right) \cdot x + \frac{1}{p(p-1)} \cdot \frac{x^2}{2} \right) 
\cdot \log_2 \e \; ,
\]
or equivalently
\[
x^2 + 2p(p-1) \ln \left( \frac{1-p}{p} \right) x - \frac{2 \theta p(p-1)}{\log_2 \e} = 0 \; ,
\]
we obtain two solutions for the intermediate $x$, namely 
\begin{eqnarray*}
x & = & \frac{1}{2} \Bigg( - 2 p(p-1) \ln \left( \frac{1-p}{p} \right) \\
& \pm & \sqrt{4 p^2 (p-1)^2 \ln^2 \left( \frac{1-p}{p} \right) 
+ \frac{8 \theta p(p-1)}{\log_2 \e}} \Bigg) \nonumber \\
& = & - p(p-1) \ln \left( \frac{1-p}{p} \right) \\
& \pm & \sqrt{ \left( p (p-1) \ln \left( \frac{1-p}{p} \right) \right)^2 
+ \frac{2 \theta p(p-1)}{\log_2 \e} } \; ; \nonumber
\end{eqnarray*}
however, only one of these solutions is positive:
\begin{eqnarray*}
x & = & - p(p-1) \ln \left( \frac{1-p}{p} \right) \\
& + & \sqrt{ \left( p (p-1) \ln \left( \frac{1-p}{p} \right) \right)^2 
+ \frac{2 \theta p(p-1)}{\log_2 \e} } \; . 
\end{eqnarray*}
The later equality can be rewritten as 
\begin{eqnarray*}
x & = & p(p-1) \ln \left( \frac{1-p}{p} \right)  \\
& \cdot & \Bigg( - 1 + \sqrt{ 1 + \frac{2 \theta}{p (p-1)
\left( \ln \left( (1-p)/p \right) \right)^2 \log_2 \e}} 
\Bigg) \; .  
\end{eqnarray*}
Using Taylor series approximation 
\[
\sqrt{1 + \chi} = 1 + \frac{1}{2} \chi - \frac{1}{8} \chi^2 + O(\chi^3) \; , 
\]
for small values of $\chi$, this becomes 
\begin{eqnarray}
x & = & p(p-1) \ln \left( \frac{1-p}{p} \right) \cdot \nonumber \\
&& \hspace{-3ex} \left( - 1 
+ 1 + \frac{\theta}{p(p-1) \left( \ln \left( (1-p)/p \right) \right)^2 \log_2 \e} \right. \nonumber \\
&& \hspace{-3ex} \left. - \frac{1}{2} \cdot \frac{\theta^2}{p^2(p-1)^2 \left( \ln \left( (1-p)/p \right) \right)^4 (\log_2 \e)^2} + O(\theta^3) \right)  \nonumber \\
& = & \frac{\theta}{\log_2 \left( (1-p)/p \right)} \nonumber \\
&& \quad - \frac{1}{2} \cdot
\frac{\theta^2 \log_2 \e}{p(p-1) \left( \log_2 \left( (1-p)/p \right) \right)^3} + O(\theta^3) \; . 
\label{eq:value-x}
\end{eqnarray}
We substitute the evaluation of value of $x$ in~(\ref{eq:value-x}) into the equation~(\ref{eq:entropy-taylor-1}).
Thus, we obtain
\begin{eqnarray}
&& \entropy_2^{-1} \left( \entropy_2(p) + \theta + O(\theta^3) \right) = 
p + \frac{\theta}{\log_2 \left( (1-p)/p \right)} \nonumber \\
&& \quad - \frac{1}{2} \cdot
\frac{\theta^2 \log_2 \e}{p(p-1) \left( \log_2 \left( (1-p)/p \right) \right)^3} + O(\theta^3)   \; . 
\label{eq:almost-final}
\end{eqnarray}
If $p < \half$ is fixed and $\theta$ is small, then the value of $\entropy_2(p) + \theta$ is bounded away 
from $1$. In this case, the derivative of $\entropy_2^{-1}(\chi)$ at point 
$\chi = \entropy_2(p) + \theta $ is bounded, and, therefore 
\[
\entropy_2^{-1} \left( \entropy_2(p) + \theta + O(\theta^3) \right)
= \entropy_2^{-1} \left( \entropy_2(p) + \theta \right)  + O(\theta^3) \; . 
\]
Then, the equality~(\ref{eq:almost-final}) becomes
\begin{eqnarray*}
\entropy_2^{-1} \left( \entropy_2(p) + \theta \right) = 
p + \frac{\theta}{\log_2 \left( (1-p)/p \right)} \\
- \frac{1}{2} \cdot
\frac{\theta^2 \log_2 \e}{p(p-1) \left( \log_2 \left( (1-p)/p \right) \right)^3} + O(\theta^3)   \; . 
\end{eqnarray*} 
Finally, we substitute $\theta = \varepsilon ( 1 - \entropy_2(p))$ and receive that
\begin{eqnarray*}
\entropy_2^{-1} \left( \entropy_2(p) + \varepsilon ( 1 - \entropy_2(p)) \right) = 
p + \frac{ \varepsilon ( 1 - \entropy_2(p))}{\log_2 \left( (1-p)/p \right)} \\
- \frac{1}{2} \cdot
\frac{ \varepsilon^2 ( 1 - \entropy_2(p))^2 \log_2 \e}{p(p-1) \left( \log_2 \left( (1-p)/p \right) \right)^3}  + O(\varepsilon^3)  \; ,
\end{eqnarray*}
thus completing the proof of the lemma. $\qed$

\section*{Acknowledgment}
The authors are thankful to Ronny Roth and Tom Richardson 
for several helpful suggestions. The authors would also like 
to thank the anonymous Reviewer C and Amin Shokrollahi for
helpful comments that substantially improved the manuscript. 
The support of DIMACS is gratefully acknowledged.


\end{document}